\documentclass[%
reprint,
showpacs,preprintnumbers,
amsmath,amssymb,
aps,
prb,
]{revtex4-1}

\usepackage{graphicx}
\usepackage{dcolumn}
\usepackage{bm}

\begin{document}


\title{The radial breathing mode in CNT - the nonlinear theory of the resonant energy exchange}

\author{V.V. Smirnov$^1$}%
\email{vvs@polymer.chph.ras.ru}
\author{L.I. Manevitch$^1$}
\author{M.Strozzi$^2$}
\author{F.Pellicano$^2$}
\affiliation{%
 $^1$Institute of Chemical Physics, RAS,
4 Kosygin str., 119991 Moscow, Russia
$^2$ Department of Engineering "Enzo Ferrari", University of Modena and Reggio Emilia, Via Pietro Vivarelli 10/1, 41125 Modena, Italy}%

\date{\today}

\begin{abstract}
We demonstrate the new specific phenomenon of the long-time resonant energy exchange in the carbon nanotubes (CNTs) with the radial breathing mode as a carrier. 
This process turns out to be stable in the wide range of the excitation energy.
It is shown that the modified nonlinear Schr\"odinger equation, obtained in the framework of nonlinear elastic thin shell theory, describes the CNT nonlinear dynamics  in the considered  frequency band.
The numerical integration of the  thin shell theory equations confirms the results of the analytical study.
\end{abstract}

\maketitle



\section{Introduction}\label{Int}

The study of nonlinear dynamics of the carbon nanotubes (CNTs) is neccessary to describe the large elastic deformation of the CNTs \citep{Soltani2012, Mahdavi2011}, the processes of the phonon-phonon interactions \citep{Gambetta2006, Greaney2009, DeMartino2009, Greaney2007, Eichler2012} and the thermal properties of the nanotubes \citep{BLi2005, Savin2009, NanoLett12}. 
The data of the numerical (computer simulation) and \textit{in situ} experiments may be compared and understood in the framework of unified viewpoint only taking into account the realistic potentials of the interatomic interaction \citep{Rafiee2014}. 
In addition to the physical nonlinearity (which reflects the nonlinear character of the interatomic forces), the nonlinearity of the geometric origin should be allowed for the framework of commonly used continuum approach to the CNT mechanics (the elastic thin shell theory, the beam theory or the finite element method) \citep{Hua2012}.
Besides the obvious consequncies of nonlinearity, such as a manifestation of the nonlinear elasticity or the loss of the stability under the large deformation \citep{Yakobson96, Chen2006, Jiang2010}, there are some nonlinear effects which associate the CNT with the wide class of the nonlinear one-dimensional lattices \citep{Scott}.
It means that the effects of anomalous diffusion and the heat conductivity may be occured in the CNTs \citep{BLi2005}.
The phenomenon of the slow energy exchange and the capture of the vibrations in some domain of the CNT may appear in the CNT vibrations \citep{VVS2010, VVS2014}.


The radial breathing mode (RBM) belongs to the most well known Raman-active oscillation branch of the carbon nanotubes (CNT) vibrations and there are many studies of the RBM, which are based on the different approaches - since the continuum shell theory up to the quantum \textit{ab initio} methods \citep{Chico06, Chang2007, NewJPhys2003, Lawler2005}.
The measurement of the RBM frequency allows to  identify the CNT raduis  \citep{Rao97, Saito1998, Dresselhaus00} as well as to determine the critical pressure of the structural transition \citep{Yang2007, Lebedkin2006}.
The role of radial breathing vibrations for  the transport properties of water molecules across SWCNTs has been also  discussed recently \citep{QZhang2013}. 
One should note that the most of the CNT dynamics studies deal with the molecular dynamics (MD) simulation or the continuum aproach (the elastic thin shell or beam theory).
However, because of the extreme complexity of the phonon band structure the dynamics of the different types of the CNT vibartions can not be described in the framework of the single approach.
In particular, the reduced nonlinear theory of the elastic thin shell under appropriate physical hypothesis is required to study the low-frequency circumferential flexure modes \citep{VVS2014, SoundVibr2014}.
 To investigate analytically the nonlinear effects, which may occur in the radial breathig branch of the CNT vibration, a new model in the framework of the thin elastic shell theory is needed.

However, one should remember that the RBM belongs to the optical branch and the specific approach is required to analyse the nonlinear effects in the long wave domain, in view of the eigenfrequencies crowding and, as a sequency, a possibility of the resonant interaction of the nonlinear normal modes (NNMs).
The adequate study of such highly non-stationary resonance processes is impossible in the framework of the NNM concept because of   the closeness of eigen frequencies that leads to appearance of a slow time scale and  to the strong mutual interaction of the resonating NNMs.
The key event in the stationary dynamics of  the systems like the Fermi-Pasta-Ulam, sine-Gordon lattices is instability of one of resonating NNMs, which can be treated as an origin of the vibration localization \citep{Daumonty1997, Zakharov2009}.
It was shown recently \citep{VVS2010} that there is a need an alternative approach to describe the highly non-stationary resonance processes. 
This approach is based on the new dynamical notion which is connected with the concept of Limiting Phase Trajectory (LPT) describing the maximum possible energy exchange between different domains of the nonlinear system. 
The NNMs  can be considered as the "ground states" in the slow time, and the motions,  which are generated by their resonance  interaction are new elementary excitations.
Such consideration is the convinient when the initial conditions provide the excitation only the resrict set of the NNMs with close frequencies (the low temperatutes and/or the selective initiations of the system oscillations).
The most intuitive example of such system is a pair of weakly coupled nonlinear identical oscillators \citep{Man07}.
Instead of conventional transition to modal presentation (in the terms of in-phase and out-of-phase NNMs) the description in the terms of the oscillators themselves  turns out to be more adequate, because on the contrary to the resonating modes they are weakly interacting.
As it was shown early, similar behaviour is  the specific one in the quasi-one-dimensional nonlinear extensive systems (nonlinear lattices \citep{VVS2010, DAN2010, VVS2011} and CNTs \citep{VVS2014}).
Some domains of the mentioned systems may be singled out, the behaviour of which turns out to be similar the nonlinear oscillators and the energy may be exchanged between them or may be localized in one of them.

In this paper we consider the resonant interaction of the nonlinear normal modes (NNMs) near the left (low-frequency) edge of the radial breathing branch in the framework of the nonlinear dynamical equation for the radial component of the displacement field. 
This equation was obtained from the nonlinear Sanders-Koiter thin shell theory in the asymptotic limit under hypotesis of the smalness of the Poisson ratio and it likes the nonlinear Schr{\"o}denger equation with the specific type of the nonlinearity. 
The spectra of the RBM branch for the CNTs with various aspect ratios and under different bondary conditions are compared with those obtained by the direct numerical integration of the equations of the Sanders-Koiter thin shell theory. 
The dynamical regime of the intensive energy exchange results from the nonlinear normal modes interaction. 
The direct numerical integration of the nonlinear equations of the Sanders-Koiter theory confirm the estimations following from the analytical model.


\section{The model}\label{model}

The long wave length dynamics of carbon nanotubes is the obvious area of applicability of the classical theory of thin elastic shell, because  the restrictions resulting from the plastic deformation are absent at the nanoscale level.
The only complicating factor is the uncertainty of the parameter characterizing the thickness of the CNT \citep{Huang06}.
However, this parameter does not play any significant role in the description of the axisymmetric deformations like the radial breathing vibrations.
In such a case the momentless thin shell theory is a well approximation for the description of the long wave length  RBMs. 
As it will be shown further, even the nonlinear effects don't change this conclusion.
The applicability of a well-designed thin elastic shell theory allows us to obtain an effective description of the vibrational spectrum in the framework of the linear approximation. 

\begin{figure}
\centering{
 \includegraphics[width=70 mm]{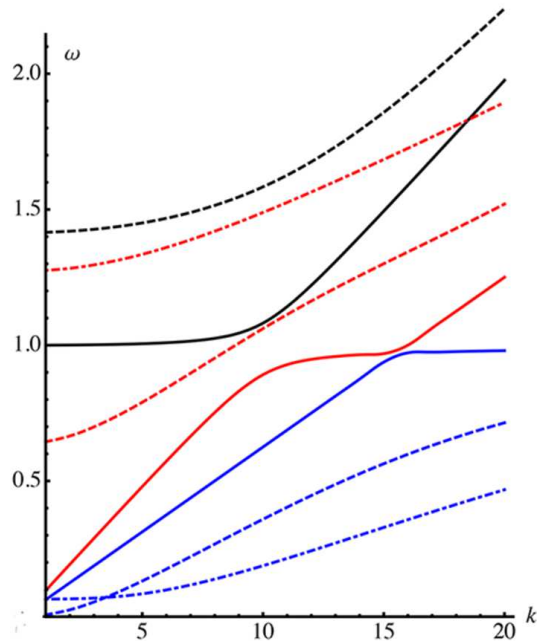}
}
\caption{
The CNT spectrum according to the exact Sanders-Koiter thin shell theory: solid curves correspond to circumferential wave number $n=0$, dashed ones - to $n=1$ and dot-dashed one - to $n=2$. 
The insert shows the small wave number part of the CFM branch.
 All the frequencies $\omega$ are measured in dimensionless units and  $k$ - denotes the number of longitudinal half-waves along the CNT.}
\label{fig:CNT2chain}
\end{figure}
It is convenient to use the dimensionless variables which determine the elastic deformation of thin circular  shell. 
In such a case all components of the displacement field ($u$ - longitudinal, along the CNT axis, $v$ - tangential and $w$ - radial, respectively) are measured in the units of the CNT radius $R$. 
The displacements and respective deformations refer to the middle surface of the shell. 
The coordinate along the CNT axis  $\xi=x/L$ is measured via the length of nanotube and varies from $0$ up to $1$, and $\varphi$ is the azimuthal angle. 
One can define the dimensionless energy and time  variables, which are measured in the units $E_{0}=YRLh/(1-\nu^{2})$ and $t_{0}=1/\sqrt{Y/\rho R^{2}(1-\nu^{2})}$, respectively. 
Here $Y$ is the Young modulus of graphene sheet, $\rho$ - its mass density, $\nu$ - the Poisson ratio of CNT, and $h$ is the effective thickness of CNT wall. 
There are two dimensionless geometric parameters which characterize CNT: the first of them is inverse aspect ratio $\alpha=R/L$ and the second - effective thickness shell $\beta=h/R$.

The energy of elastic deformation of the CNT in the dimensionless units is written as follows:

\begin{equation}\label{energy_elst2}
\begin{split}
E_{el} =\frac{1}{2} \int\limits_{0}^{1} \int\limits_{0}^{2 \pi}\left(\varepsilon_{\xi}^{2} + \varepsilon_{\varphi}^{2} + 2 \nu \varepsilon_{\xi} \varepsilon_{\varphi} + \frac{1 - \nu}{2} \varepsilon_{\xi \varphi}^{2}\right)d\xi d\varphi +\\
+\frac{\beta^{2}}{24} \int\limits_{0}^{1} \int\limits_{0}^{2 \pi}\left(\kappa_{\xi}^{2} + \kappa_{\varphi}^{2} + 2 \nu \kappa_{\xi} \kappa_{\varphi} + \frac{1 - \nu}{2} \kappa_{\xi \varphi}^{2}\right)d\xi d\varphi ,
\end{split}
\end{equation}

where $\varepsilon_{\xi}$, $\varepsilon_{\varphi}$ and $\varepsilon_{\xi \varphi}$ are the longitudinal, circumferential and shear deformations, and $\kappa_{\xi}$, $\kappa_{\varphi}$ and $\kappa_{\xi \varphi}$ are the longitudinal and circumferential curvatures, and torsion, respectively: 

\begin{equation}\label{deformation}
\begin{split}
\varepsilon_{\xi} = \alpha \frac{\partial u}{\partial \xi} + \frac{\alpha^{2}}{2}( \frac{\partial w}{\partial \xi})^{2} +\frac{1}{8}(\alpha \frac{\partial v}{\partial \xi}-\frac{\partial u}{\partial \varphi})^{2} \\ 
\varepsilon_{\varphi} = \frac{\partial v}{\partial \varphi} + w + \frac{1}{2} (\frac{\partial w}{\partial \varphi} -v)^{2}+\frac{1}{8}(\frac{\partial u}{\partial \varphi}-\alpha \frac{\partial v}{\partial \xi})^{2}  \\
 \varepsilon_{\xi \varphi} = \frac{\partial u}{\partial \varphi} + \alpha \frac{\partial v}{\partial \xi} +\alpha \frac{\partial w}{\partial \xi}(\frac{\partial w}{\partial \varphi}-v) 
 \end{split}
\end{equation}

\begin{equation}\label{curvation}
\begin{split}
\kappa_{\xi} = - \alpha^{2}  \frac{\partial^{2} w}{\partial \xi^{2}}, \quad  
\kappa_{\varphi} =  \left(\frac{\partial v}{\partial \varphi} - \frac{\partial^{2} w}{\partial \varphi^{2}} \right), \\ 
\kappa_{\xi \varphi} =  \left( - 2 \alpha \frac{\partial^{2} w}{\partial \xi \partial \varphi} + \frac{3 \alpha}{2} \frac{\partial v}{\partial \xi} - \frac{1}{2} \frac{\partial u}{\partial \varphi}\right).
\end{split}
\end{equation}

\begin{figure}
 \includegraphics[width=70mm]{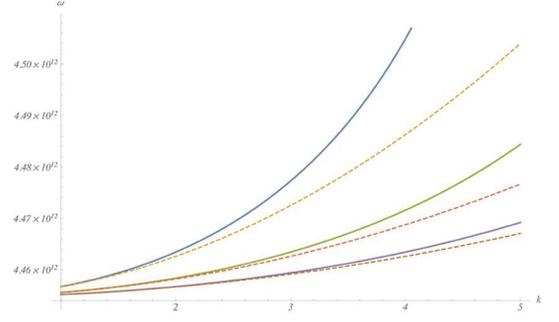}
 \caption{(Color online) The RBM spectra for the CNTs with different aspect ratios according to eq. \eqref{eq:eigenvalue} (solid lines) and eq. \eqref{eq:eigenvalue2} (dashed lines). The aspect ratios are equal to 20 (blue and dashed yellow), 30 (green and dashed orange), and 40 (violet and light braun).}
 \label{fig:RBM_spectrum}
 \end{figure}

Taking into account that the radial breathing normal modes do not depend on the azimuthal angle $\varphi$ (the respective azimuthal wave number $n=0$) and the transversal displacement $v=0$, one can write the equation of  motion as follows:

\begin{multline}\label{eq:full}
\frac{\partial^2 u}{\partial \tau^2}  - \alpha \nu \frac{\partial w}{\partial \xi} - \alpha^{2} \frac{\partial ^{2} u}{\partial \xi^{2}} - \alpha^{3} \frac{\partial w}{\partial \xi} \frac{\partial ^{2}w}{\partial \xi^{2}}= 0 \\
\frac{\partial^{2}w}{\partial \tau^{2}} + w + \alpha \nu \frac{\partial u}{\partial \xi}- \alpha ^{2} \nu \left( \frac{1}{2}  \left( \frac{\partial w}{\partial \xi} \right)^2 + w \frac{\partial^{2} w}{\partial \xi^{2}} \right) \\ -\alpha ^3 \left(  \frac{\partial^{2} u}{\partial \xi^{2}} \frac{\partial w}{\partial \xi} + \frac{\partial  u}{\partial \xi} \frac{\partial ^{2} w}{\partial \xi^{2}} \right)+   \\   \alpha ^{4} \left( \frac{1}{12}  \beta ^{2} \frac{\partial ^{4}w}{\partial \xi^{4}}-\frac{3}{2}   \left( \frac{\partial w}{\partial \xi} \right)^{2} \frac{\partial ^{2}w}{\partial \xi^{2}} \right) = 0
\end{multline}
Dispersive curves of the linearized problem contains two branches:

\begin{equation}\label{eq:eigenvalue}
\omega^{2}=\frac{1}{2} \left(1+\alpha^{2} k^{2} \pm \sqrt{(1-\alpha k)^{2} +4 \alpha^{2} \nu^{2} k^{2}} \right)
\end{equation}

the one of which corresponds to the longitudinal acoustic modes (sign ''+'') and the second one - to the radial breathing modes (sign ''-''). $k$ is the longitudinal wave number.
Figure \ref{fig:RBM_spectrum} shows the eigenfrequencies of the long wavelength modes for the CNTs with different aspect ratios.
In spite of that the real wave number for the periodic boundary conditions starts from the value $k=1$, the product $\alpha k$ is small if the CNT is long enough. 
The long wavelength limit of \eqref{eq:eigenvalue} shows the spectrun crowding:
\begin{equation}\label{eq:eigenvalue2}
\omega \simeq 1+\frac{1}{2}\alpha^2 \nu^2 k^2
\end{equation}

The respective eigenvector
\begin{equation}\label{eq:eigenvector}
\left( u,w\right)=\left( -\alpha \nu k,1 \right)
\end{equation}
 shows the relationship between longitudinal and radial components of the displacement field.
 
 Taking into account the expression \eqref{eq:eigenvector} one can find the relation between $u$ and $w$ in the coordinate space as
 
 \begin{equation}\label{eq:ucomp}
 u(\xi,\tau) = -\alpha \nu \frac{\partial w(\xi,\tau)}{\partial \xi}.
 \end{equation}
 
Using this ratio one can rewrite second equation of  \eqref{eq:full} as follows:

\begin{multline}\label{eq:weq}
\frac{\partial ^{2} w}{\partial \tau^{2}} + w^{2} -\alpha^2 \nu^2 \frac{\partial ^2 w}{\partial \xi^2}-\alpha ^2 \nu \left( \frac{1}{2}  \left( \frac{\partial w}{\partial \xi} \right)^2 + w \frac{\partial ^{2}w}{\partial \xi^{2}} \right)  \\ +\alpha^{4} \left( \frac{1}{12} \beta ^2 \frac{\partial ^{4} w}{\partial \xi^{4}} + \nu \frac{\partial}{\partial \xi} \left( \frac{\partial w}{\partial \xi} \frac{\partial ^2 w}{\partial \xi^2} \right)-\frac{3}{2} \left( \frac{\partial w}{\partial \xi} \right)^2 \frac{\partial^{2}w}{\partial \xi ^2} \right) \\ =0
\end{multline}

To perform the asymptotic analysis of the long wavelength dynamics of the RB modes, it is convenient to rewrite  equation \eqref{eq:weq} using complex variables:

\begin{equation*}\label{eq:complex1}
\psi=\frac{1}{\sqrt{2}} \left( \frac{\partial w}{\partial \tau}+ i w \right).
\end{equation*}

Returning to expression \eqref{eq:eigenvector} one can see that the assumption of the Poisson ratio smallness  $\nu < 1$ allows to use it as a small parameter. 
Using the multiple time scales $\tau_{0}=\tau$, $\tau_{1}=\nu \tau$, $\tau_{2}=\nu^{2} \tau$, etc., and expanding the function $\psi$ into series of the small parameter $\nu$: 

\begin{equation*}
\psi=\nu \left( \psi_{0}+\nu \psi_{1} + \nu^{2} \psi_{2}+\dots \right) e^{- i \tau},
\end{equation*}

one can get the equation for the main order amplitude in the ''slow'' time $\tau_{2}$ (see, for example, \citep{VVS2010, CFM2014} for details): 

\begin{equation}\label{eq:NLSE}
i \frac{\partial \psi_{0}}{\partial \tau_{2} } - \alpha^2 \frac{1}{2}   \frac{\partial ^{2} \psi_{0} }{\partial \xi^{2}}  - \alpha^{4} \frac{3}{8} \frac{\partial}{\partial \xi} \left( \left|  \frac{\partial \psi_{0}}{\partial \xi} \right| ^{2}  \frac{\partial \psi_{0}}{\partial \xi} \right) =0.
\end{equation}

Equation \eqref{eq:NLSE} admits the plane-wave solution

 \begin{equation}\label{eq:planewave}
 \psi_{0}= A \exp{(-i \left( \tilde{\omega} \tau_{2} - k \xi) \right)}.
 \end{equation}
 
Taking into account the ''slowness'' of the time $\tau_{2}$, the respective dispersion ratio should be written as follows:
\begin{equation}\label{eq:NLSE_dispersion}
\omega=1+\nu^{2} \tilde{\omega} = 1+\alpha^{2} \nu^{2} k^{2}+ \frac{3}{8} \alpha^{4}\nu^{2}  k^{4} A^{2}.
\end{equation}

One can see that the amplitude-independent part of \eqref{eq:NLSE_dispersion} is in accordance with the relation \eqref{eq:eigenvalue2}, while the effective positiveness of nonlinear addition points out the hard type of nonlinearity. 

Equation \eqref{eq:NLSE} is the modified Nonlinear Schr{\"o}dinger Equation (NLSE) with the gradient type of the nonlinearity.
As it is well known, the standard NLSE admits the localized solution - the envelope soliton or the breather. 
However, a possibility any localized solutions in equation \eqref{eq:NLSE} is unknown.
We will try to examine the possibility of energy localization while dealing equation \eqref{eq:NLSE}.
First of all we replace equation \eqref{eq:NLSE} by its modal representation, taking into accout only two resonant NNMs  with the wave numbers $k_{1}$ and $k_{2}$. 

\begin{equation}\label{eq:NLSE2waves}
\psi_{0}=\chi_{1}(\tau_{2}) \sin{(\pi k_{1} \xi)}+\chi_{2}(\tau_{2}) \sin{(\pi k_{2} \xi)}
\end{equation}


After substitution of solution \eqref{eq:NLSE2waves} into equation \eqref{eq:NLSE} one should use the Galerkin procedure to obtain the equations for complex amplitudes $\chi_{1}$ and $\chi_{2}$:
\begin{equation}\label{eq:2waves_1}
\begin{split}
i \frac{\partial \chi_{1}}{\partial \tau_{2}} + \delta \omega_{1} \chi_{1} -\frac{3 \sigma_{11}}{2}  \left|\chi_{1} \right|^{2} \chi _{1}  \qquad \qquad \qquad \quad  \\  
-\sigma_{12} \left( 2\left| \chi_{2} \right| ^{2} \chi _{1} + \chi_{2}^{2} \chi _{1}^{*} \right) = 0 \\
i \frac{\partial \chi_{2}}{\partial \tau_{2}} + \delta \omega_{2} \chi_{2} - \frac{3  \sigma_{22} }{2} \left|\chi_{2} \right|^{2} \chi _{2}  \qquad \qquad \qquad \quad  \\
-\sigma_{12} \left( 2 \left|\chi_{1} \right|^{2} \chi _{2} +  \chi_{1}^{2} \chi _{2}^{*} \right) = 0,
 \end{split}
 \end{equation}
 where $ \delta \omega_{i} = \frac{1}{2} \pi^{2} k_{i}^{2}$, ($ i=1,2$)  are the modal  frequency shifts (in the ''slow'' time scale $\tau_{2}$)  from the  boundary frequency $\omega_{0}=1$ of the considered brunch  and
 
 \begin{equation}\label{eq:sigma}
 \sigma_{ij} = \frac{3}{16} \alpha^{4}  k_{i}^{2} k_{j}^{2}  \quad (i,j=1,2).
 \end{equation}

One can see that the nonlinear terms in equations \eqref{eq:2waves_1} are separated into two groups: the terms $|\chi_{j}|^{2} \chi_{i}$  $(i,j = 1,2)$ determine the nonlinear frequency shift, while the terms  $\chi_{i}^{2} \chi_{j}^{*} ( i \ne j) $ describe the nonlinear interaction between modes. 
The Hamiltonian corresponding to equations \eqref{eq:2waves_1} can be written as 

\begin{equation}\label{eq:NLSEhamiltonian2}
\begin{split}
H= \delta \omega _1\left|\chi _{1} \right|^2+\delta \omega _2\left| \chi _{2} \right|^2  - \frac{3}{4} \left( \sigma_{11} \left| \chi _{1} \right|^4 + \sigma_{22} \left| \chi _{2} \right|^{4} \right)  \\
-\sigma_{12}\left( 2 \left| \chi _{1} \right|^2 \left| \chi _{2} \right|^{2}+\frac{1}{2} \left( \chi _{1}^{2} \chi _{2}^{*2} +\chi _{1}^{*2} \chi _{2}^{2} \right) \right) \\
\end{split}
\end{equation}

Besides the obvious energy integral \eqref{eq:NLSEhamiltonian2}, equations \eqref{eq:2waves_1}  possess another integral 

\begin{equation}\label{Occupation}
X=\left| \chi_{1} \right|^{2}+ \left| \chi_{2} \right|^{2},
\end{equation}

which characterises the excitation level of the system.
This is an analogue of the occupation number integral in quantum-mechanical terminology.

 
 As it was noticed in \citep{VVS2010, CISM2010, VVS2014, CFM2014}, the modal description of nonlinear dynamics turns out to be inconsistent under resonant conditions. 
 Therefore we need in the introduction of new weakly interacting variables
We have shown early \citep{VVS2010} that they are the linear combination of the resonating modes:

\begin{equation}\label{psi}
\phi_{1}=\frac{1}{\sqrt{2}}(\chi_{1}+\chi_{2}); \phi_{2}=\frac{1}{\sqrt{2}}(\chi_{1}-\chi_{2}).
\end{equation}
 
The new variables describe the dynamics of some parts of the CNT \citep{VVS2014} (or some groups of the particles in the effective discrete one-dimensional chain   \citep{VVS2010, DAN2010, VVS2011}), because one can see  a predominant energy concentration in certain domain of the CNT, while the other part of CNT has a lower  energy. 
Due to small difference between the modal frequencies, the selected  parts of CNT demonstrate a coherent behavior.
They are similar to two weakly coupled nonlinear oscillators and their temporal behavior looks very like to the beating in  such system.
Therefore we can consider these regions as new large-scale elementary blocks, which can be identified as unique elements of the system - the ''effective particles'' \citep{VVS2010}. 
 
  One can notice that the introduction of effective particles \eqref{psi} makes equations \eqref{eq:2waves_1} to be more complicated. 
  However, due to the presence of the integral \eqref{Occupation}, the dimensionality of the phase space of the system can be reduced. 
  The ''occupation number'' $X$ parametrizes the total excitation of the system, but the distribution of the energy is determined by the amplitudes of the ''effective particles'' as well as by the phase shift between them. 
  Actually, taking into account expression \eqref{Occupation}, one can describe the behavior of the ''effective particles'' with two real functions:

\begin{equation}\label{Angle1}
\phi_{1}=\sqrt{X} \cos{\theta}e^{-i\Delta /2} ;  \quad
\phi_{2}=\sqrt{X} \sin{\theta}e^{  i\Delta /2},
\end{equation}

where the variable $\theta$ characterizes the relative amplitudes of the ''effective particle'' and the variable $\Delta$ - the phase shift between them.

Substituting  relationships (\ref{psi},  \ref{Angle1}) into equations \eqref{eq:2waves_1}, we obtain the equations of motion in the terms of "angular" variables ($\theta, \Delta$):

\begin{equation}\label{eq:NLSEangle}
\begin{split}
\sin{2 \theta } ( \frac{\partial \theta}{\partial \tau_{2}} - \frac{1}{2} ( ( \delta \omega _{2} -\delta \omega _{1} )     - \frac{3}{4} X ( \sigma _{11}  - \sigma _{22})) \sin{ \Delta}  \\ - \frac{1}{16} X (3  \sigma _{11} -4 \sigma_{12}+ 3 \sigma_{22} )\sin{2 \Delta } \sin{2 \theta } )    = 0    \\
\sin{2 \theta} \frac{\partial \Delta}{\partial \tau_{2}}+( ( \left(\delta\omega_{1}-\delta\omega_{2} \right)+\frac{3}{4}X \left(\sigma_{22}-\sigma_{11} \right) ) \cos{\Delta}  - \\ \frac{1}{4} X (  (3 \sigma_{11} -2 \sigma_{12}+3 \sigma_{22}) \cos^{2}{\Delta} -4 \sigma_{12} ) \sin{2 \theta} )\cos{2 \theta} = 0
\end{split}
\end{equation}

\begin{figure}
 \centering{
  \includegraphics[ width=70mm]{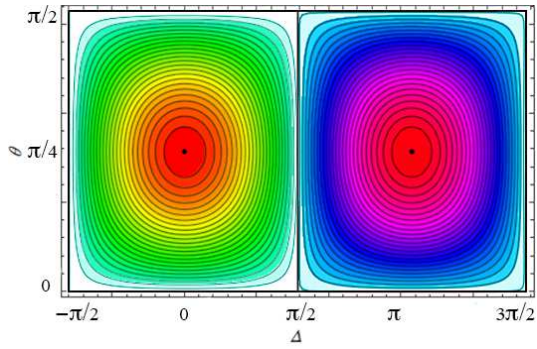}  \quad
}
\caption{(Color online) Phase portraits of the system \eqref{eq:NLSEangle} for the aspect ratio $L/R=80$ and the ''occupation number'' $X=2.5$. The LPTs mark out by the thick black lines.  }
\label{fig:NLSE_phaseportrait}
\end{figure}

There are two types of the fundamental solutions on the presented phase plane. The stationary points corresponding (in the slow time) to NNMs determine the stationary dynamics of the system. 
However, other type of phase trajectories is significant for understanding and description of highly non-stationary resonant dynamics.  
These are the trajectories, which separate the NNMs attraction domains and they are the most distant from the stationary points. 
Such trajectories correspond to the extremely non-uniform distributions of the energy (from a possible ones). 
They were classified as the Limiting Phase Trajectories (LPTs). 
The main distinction between the LPT and the separatrix is that the period of the motion along them is, generally speaking the finite  one.  
The motion along the LPT between the states $\theta = 0$ and $\theta = \pi/2$ leads to the redistribution of the energy between the different parts of the system, i.e., between the "effective particles". 
This process is an analogue of the beating in the system of weakly coupled oscillators. 
It was shown that an adequate temporal description of LPT can be obtained in the terms of non-smooth functions which are saw-tooth function and its derivative in the sense of the distributions theory.
 All of these peculiarities of the phase space of the system \eqref{eq:NLSEangle} are shown in Fig. \ref{fig:NLSE_phaseportrait}.
 The representative domains of the phase space are bounded by the intervals $0 \le \theta \le \pi/2$ and $-\pi/2 \le \Delta \le 3 \pi/2$.
 
\begin{figure}
a \includegraphics[width=35mm]{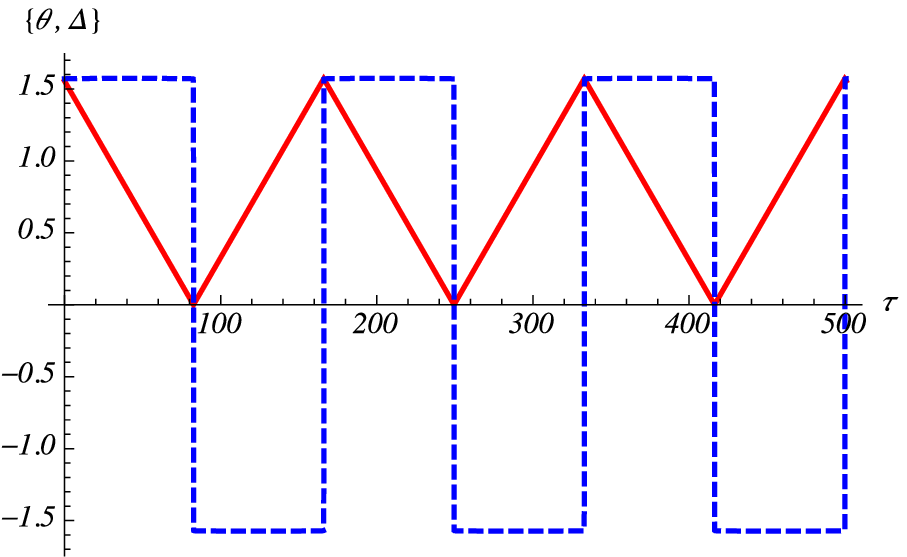} b \includegraphics[width=35mm]{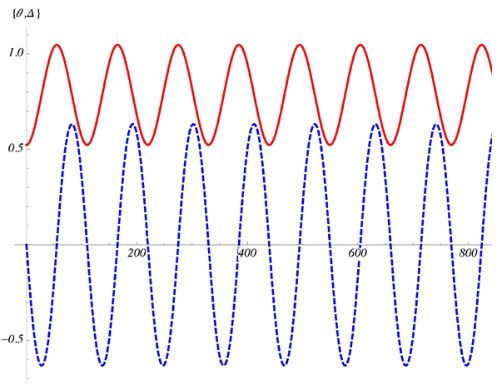}
\caption{(Color online) Time evolution of the variables $\theta$ (red solid lines) and $\Delta$ (blue dashed lines) for the CNTs with different aspect ratios: (a) $L/R=20$; (b) $L/R=80$. The ''occupation number'' $X=0.5$. One should pay the attention that $\tau$ is the ''slow'' time and the real times (in the own period of the RBM) result as the value of $\tau$ divided by the square of the small parameter.}
\label{fig:TD_curves0}
\end{figure}

The numerical solutions of Eq. \eqref{eq:NLSEangle} with the initial conditions corresponding to the immovable point ($\theta=\pi/2$, $\Delta=\pi/2$) for two values of the excitation $X$ are shown in the Fig. \ref{fig:TD_curves0} (a-b).

The solutions, which are shown in Fig. \ref{fig:TD_curves0}, correspond to a slow redistribution of the energy between the ''effective particles''.
If the initial conditions respect to the LPT, the energy exchange reachs the maximum of the possible amount. 
The period of such energy exchange may be estimated as the time of  passing the trajectory:
\begin{equation}\label{eq:period}
T=\oint{ d \tau_{2}}=\oint\frac{d \theta}{d \theta/ d \tau_{2}}
\end{equation}
The integral in equation \eqref{eq:period} can be estimated from the first of equations \eqref{eq:NLSEangle} taking into account that $\Delta \simeq \pm \pi/2$ on the LPT (see Fig. \ref{fig:NLSE_phaseportrait}) and the transition from $\Delta=-\pi/2$ to $\Delta=\pi/2$ takes no time:

\begin{equation}\label{eq:simperiod}
T=2 \int_{0}^{\pi/2} \frac{d \theta}{d \theta/ d \tau_{2}} \simeq \frac{2\pi }{ \left(\left(\delta \omega _2-\delta \omega _1\right)+\frac{3}{4} \left(\sigma _{22}-\sigma _{11}\right) X\right)}
\end{equation}

\begin{figure}
\includegraphics[width=80mm]{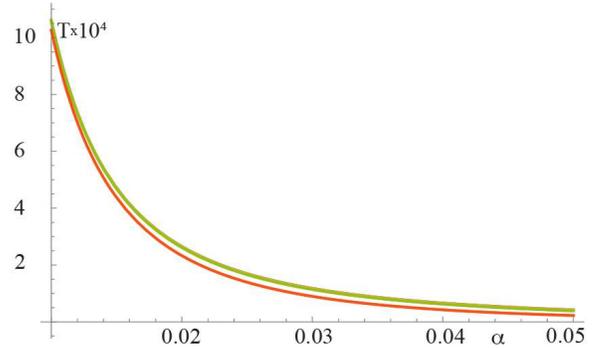}
\caption{(Color online) Period of energy exchange vs inverse aspect ratio of the CNT. Blue, orange and green curves correspond to the excitation $X=0.01,$ $0.5,$ and $1$, respectively}
\label{fig:period}
\end{figure}

The variation of the period \eqref{eq:simperiod} with the aspect ratio of the CNT is shown in Fig.\ref{fig:period}.

To analyse the possibility of the stationary states instability one should rewrite  Hamiltonian \eqref{eq:NLSEhamiltonian2} in the terms of the variables $\theta$ and $\Delta$:

\begin{multline}\label{eq:Hangle}
H(\theta, \Delta)=\frac{X}{2} \left( \left( \delta \omega _{1}+\delta \omega_{2} \right) + \left( \delta \omega_{1}- \delta \omega _{2} \right) \cos{\Delta} \sin{2 \theta} \right)  \\  +\frac{X^{2}}{16} ( 3 \sigma _{22} (1-\cos{\Delta} \sin{2 \theta} )^2   +3 \sigma _{11} (1+\cos{\Delta} \sin{2 \theta} )^2   \\  + 4 \sigma _{12} \left( 3 \cos^{2}{2 \theta}+ \sin^{2}{2 \theta}\sin^{2}{\Delta} \right) ).
\end{multline}

The conditions of instability may be formulated as:

\begin{equation}\label{eq:1bifurcation}
\frac{\partial ^{2} H}{\partial \theta^{2}} = 0
\end{equation}

at  ($\theta=\pi/4$, $\Delta=0$) or ($\theta=\pi/4$, $\Delta=\pi$).

The latter results in 

\begin{equation}\label{eq:threshold}
\begin{split}
X=\frac{2 (\delta\omega_{2}-\delta\omega_{1})}{3(\sigma_{11}-2\sigma_{12})} \quad \Delta=0;  \\
X=\frac{2 (\delta\omega_{2}-\delta\omega_{1})}{3(2\sigma_{12}-\sigma_{22})} \quad \Delta=\pi.
\end{split}
\end{equation}

Taking into account the definition of $\delta\omega_{j}$ and relationship \eqref{eq:sigma} one can see that no bifurcation at the positive values of ''occupation number'' $X$ exist.
Therefore, no localized breather-like excitatons can exist in the single walled CNT.
So, in the contrary with the low-frequency circumferential flexure vibrations \citep{VVS2014}, only the intensive energy exchange is possible in the radial breathing branch. 

The analysis  performed above is based on the asymptotic expansion of the equations in the framework of the nonlinear Sanders-Koiter elastic thin shell theory. 
Only two modes in the RB branch were taking into account.
Therefore, a verification of our conclusion by the independent numerical methods is needed.
Fig. \ref{fig:RBM_spectrum} shows that if the aspect ratio of the CNT is large enough, more than two modes can be under resonant conditons.
So, the influence of the other part of the spectrum is very important for the estimation of the reliability of the obtained results. 
Our approach consists in the direct numerical integration of the modal nonlinear equations of the Sanders-Koiter thin shell theory. 
The detail procedure was described for the circumferential flexure modes in \citep{SoundVibr2014}.
Therefore, we consider the method extremely shortly.
In order to carry out the numerical analysis of the CNT dynamics, a two-step procedure was used: i) the displacement field was expanded by using a double mixed series, then the Rayleigh-Ritz method was applied to the linearized formulation of the problem, in order to obtain an approximation of the eigenfunctions; ii) the displacement fields are re-expanded by using the linear approximated eigenfunctions, the Lagrange equations were then considered in conjunction with the nonlinear elastic strain energy to obtain a set of nonlinear ordinary differential equations of motion.

To satisfy the boundary conditions the displacement field was expanded into series 

\begin{equation}\label{eq:expansion1}
r(\xi ,\varphi,t )=\left[ \sum _{m=0}^{M_{u} }\sum _{n=0}^{N}R_{m,n} T_{m}^{*} (\xi ) \right] f(t) \\
\end{equation}

where function $r(\xi, \varphi,t)$ substitutes the displacements $u$, $v$ or $w$. 

In equations \eqref{eq:expansion1}  $T_{m}^{*} (\xi )=T_{m} (2\xi -1)$ are the Chebyshev orthogonal polynomials of the $m-th$ order and $f(t)$ describes the time evolution of the CNT vibrations.

Such an expansion allows to estimate the natural frequencies (eigenvalues) and modes of vibrations (eigenvectors) under various boundary conditions.
The results of performed calculation show that the eigenspectrum values are in the good accordance with the estimations made in the framework of reduced Sanders-Koiter theory discussed above (see Fig. \ref{fig:RBM_spectrum}).

\begin{figure*}
a)\includegraphics[width=40mm]{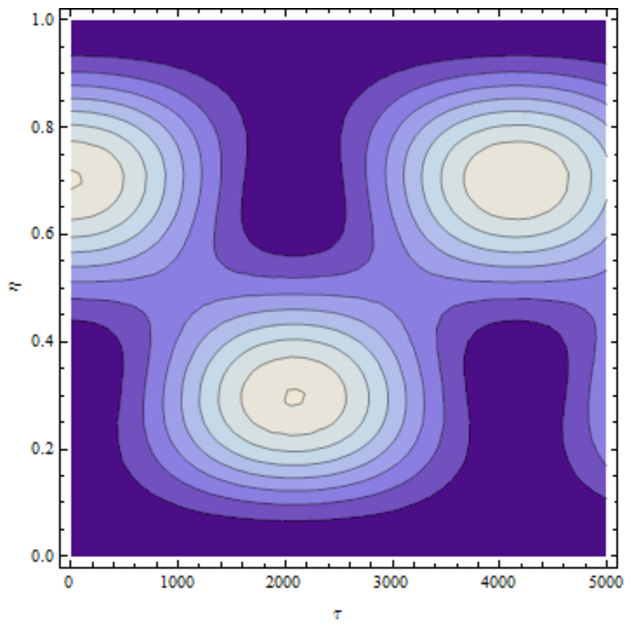} \quad
b)\includegraphics[width=40mm]{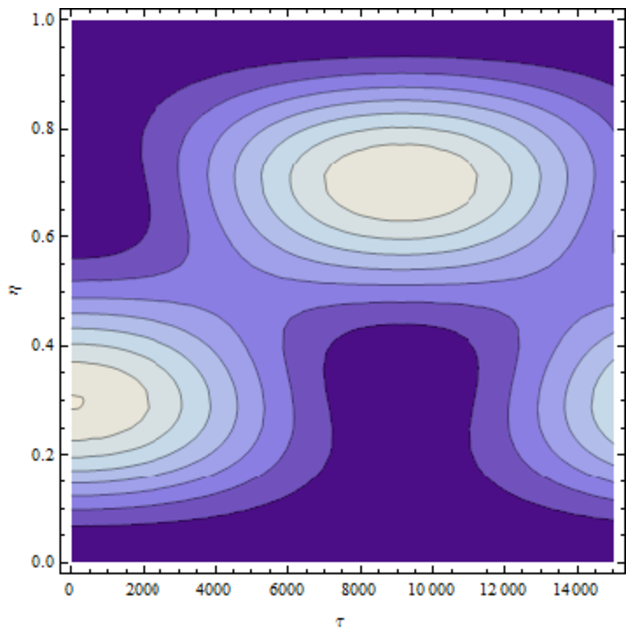} \quad
c)\includegraphics[width=40mm]{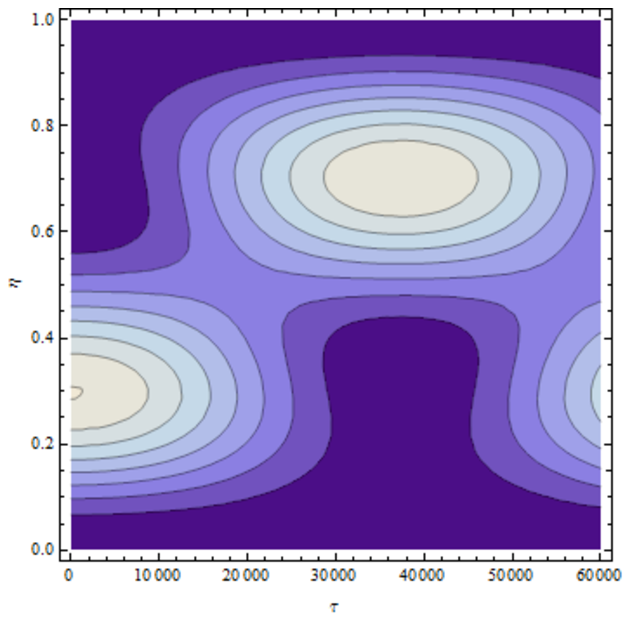}
\caption{(Color online) The energy exchange in the CNTs with different aspect ratios: (a) $L/R=20$, (b) $L/R=40$, (c) $L/R=80$. The initial ''occupation number'' $X=0.5$. The dark violet and light beige areas correspond to low and high density of the energy.}
\label{fig:beathing}
\end{figure*}

In the nonlinear analysis, the full expression of the dimensionless potential energy $E_{el}$ containing terms up to the fourth order (cubic nonlinearity), is considered.
Using the Lagrange equations

\begin{equation}\label{eq:lagr_eq}
\frac{\partial^2 r}{\partial \tau^2 } +\frac{\delta E_{el}}{\delta r } =0, \quad r=\{u, v, w\}
\end{equation}

a set of nonlinear ordinary differential equations is obtained; these equations must be completed with suitable initial conditions on displacements and velocities. This system of nonlinear equations of motion was finally solved by using the implicit Runge-Kutta numerical methods with suitable accuracy, precision and number of steps. The solution of nonlinear equations with initial conditions in the vicinity of the the LPT shows the energy exchange process, the period of which  coincides with equation \eqref{eq:simperiod}  for the wide interval of aspect ratios  and excitation amplitudes (see Fig. \ref{fig:beathing}).

\section{Conclusion}\label{conclusion}

As a summary we would like to notice that the phenomenon of the partial or intensive full energy exchange is the specific one for the resonating nonlinear normal modes. 
The generalized form of the Hamiltonian  \eqref{eq:NLSEhamiltonian2} reflects the most common sutiation for nonlinear interaction in very different systems \citep{Man07, VVS2010, DAN2010, CISM2010, VVS2014}. 
The result of this interaction is determined by the relations between the parametrs of nonlinearity ($\sigma_{ij}$) and the frequency differences ($\delta\omega_{j}$). 
The most important notion in that is the Limiting Phase Trajectory, which corresponds to the ''elementary process'' associated with the energy exchange or capture. 
One should emphasize that only the analysis performed in the framework of this conception is capable of predicting of  the main bifurcation in the dynamics of the system, in spite of that the quantitative results may be deficient due to restricted description of the system.

Concerning with the radial breathing vibrations the absence of energy capture and localization is  rather unexpected result, because the instability of low-frequency breathing modes seems intuitively evident. 
However, the presence of the strong gradient nonlinearity in equation \eqref{eq:NLSE} breaks down this viewpoint. 
It is the bright distinction of the radial breathing branch from the circumferential flexure one \citep{VVS2014}.

\begin{acknowledgments}
Two of authors (VVS and LIM) are grateful to Russia Basic Research Foundation (grant 08-03-00420a) and Russia Science Foundation (grant 14-17-00255) for the financial supporting of this work.
\end{acknowledgments}

\bibliography{RBM}

\end{document}